\documentclass[%
 aip,
 amsmath,amssymb,
reprint,%
]{revtex4-1}
\usepackage{color,soul}
\usepackage{graphicx}
\usepackage{dcolumn}
\usepackage{bm}

\usepackage[utf8]{inputenc}
\usepackage[T1]{fontenc}
\usepackage{lmodern}

\usepackage{multirow}
\usepackage[table,xcdraw]{xcolor}

\usepackage[colorlinks=true,allcolors=blue]{hyperref}
\usepackage{cancel}
\usepackage{ulem}
\usepackage{soul}

\begin{document}

\preprint{AIP/123-QED}

\title[]{Acoustic interaction force between two particles immersed in a viscoelastic fluid}
\author{Fatemeh Eslami}
\affiliation{ 
Department of Physics, K.N. Toosi University of Technology, Tehran 15875-4416, Iran.}%
\author{Hossein Hamzehpour}%
\email{ hamzehpour@kntu.ac.ir}
\affiliation{ 
Department of Physics, K.N. Toosi University of Technology, Tehran 15875-4416, Iran.}%
\author{Sanaz Derikvandi}%
\affiliation{ 
Department of Physics, K.N. Toosi University of Technology, Tehran 15875-4416, Iran.}%

\author{S. Amir Bahrani}
\email{amir.bahrani@imt-nord-europe.fr}
\affiliation{
IMT Nord Europe, Institut Mines Télécom, Univ. Lille, Center for Energy and Environment, F-59000 Lille, France.
}

\date{\today}
\begin{abstract}
The interaction acoustic radiation force in a standing plane wave applied to each small solid sphere in a two-particle system immersed in a viscoelastic fluid is studied in a framework based on perturbation theory. In this work, the first- and second-order perturbation theories are used in the governing equations with considering the upper-convected maxwell model to obtain mathematical modeling. We use the finite element method to carry out simulations and describe the behavior of the viscoelastic fluid. The mathematical development is validated from three literature case studies: a one-particle system in a viscous fluid, a two-particle system in a viscous fluid, and a one-particle system in a viscoelastic fluid. The novelty of this study is to establish the acoustic interaction force between two spherical particles immersed in a viscoelastic fluid. The results show that the acoustic interaction force between two spheres is greater in a viscous fluid in comparison with the viscoelastic fluid with the same shear viscosity. This behavior is due to the relaxation time effect. 
A mathematical formula is proposed for the acoustic interaction force between particles located close to each other in a viscoelastic fluid.

\end{abstract}
\maketitle

\textit{Introduction}. Many studies have been done on particle manipulation using acoustophoresis which is essential for analyzing micro-sized particles in numerous biological and chemical applications\cite{King1934,Crum1975,Embleton1962,Fox1940}. Acoustophoresis is the movement of particles in micro-channels by ultrasound.  Particles exposed to an external acoustic wave that is mainly a standing wave, are subjected to a time-averaged force from scattering waves called the acoustic radiation force\cite{Gorkov1962,Yosioka1955,Hasegawa1988,Mitri2005,FMitri2005,Doinikov2001, Mitri2006,Settnes2012}.  
 For a two-particle system in an acoustic field, the total radiation force applied to each particle consists of primary and secondary forces. The acoustic primary force is due to the incident and the scattered waves from the surface of the particle. The primary force pushes particles into the pressure node or antinode depending on their acoustic contrast factor\cite{Settnes2012}. The secondary (interaction) force is exerted on each particle owing to the effect of the scattered wave from the other one\cite{Doinikov1999,Silva2014,Lopes2016,Zheng1995,Doinikov2002,Sepehrirahnama2015,Sepehrirahnama2016,Pelekasis2004}. In practice, there are many micro-sized particles in biological solutions and suspensions; therefore, the interaction force between them plays a significant role and makes our study important.

At first, the acoustic interaction force was studied by Bjerknes\cite{Crum1975} to investigate the acoustic inter--particle force between a pair of bubbles, based on the theory of rescattering of scattered waves. Then, Apfel and Embleton\cite{Embleton1962} obtained an approximation for the acoustic interaction force between two spheres using King's and Yosioka's methods \cite{King1934, Yosioka1955}. Doinikov and Zavtrak\cite{Zavtrak1995,Zavtrak1997} applied the multipole re--expansion method. They used monopole and dipole terms of the multipole series expansion to calculate the primary and interaction forces. Doinikov\cite{Doinikov2001} used five terms of this expansion to estimate the secondary force between two bubbles in the water. Bruus and Silva\cite{SivaBruus2014} in $2014$ indicated that the interaction force can be derived from the potential field just like the primary force. Sepehrirahnama and his colleagues \cite{Sepehrirahnama2016} in $2016$ represented the effect of viscosity and acoustic streaming on the inter-particle radiation force between two rigid spheres and they proposed a numerical algorithm for calculation of the primary and total radiation forces. Their method had no restriction on the size of the spheres because of the use of higher-order multipole terms.

Although particle movement in Newtonian fluids has been widely studied so far \cite{Doinikov2001, Sepehrirahnama2015}, there is little research on particle movement in non-Newtonian fluids. Non-Newtonian fluids such as blood, suspensions, and many other fluids are very ubiquitous in our daily life. Dual et al.\cite{Dual2021} in 2021 studied the acoustic radiation force on a solid particle in an acoustically excited  viscoelastic fluid. They described the fluid motion by the compressible Oldroyd-B model and no restrictions were imposed on the particle size with respect to the acoustic wavelength and the viscous penetration depth in their method. 

Another second-order nonlinear effect that is exerted by the incident acoustic wave on the particles is called acoustic streaming\cite{Amir2020}, but in this contribution, our focus is on the acoustic radiation force effect.
To our knowledge, there is no study on acoustic interaction force between spheres suspended in a viscoelastic fluid.  Therefore it is important to invest our research focus on calculating the acoustic interaction force in non-Newtonian fluids to develop our understanding of particles' behavior inside such fluids.

The novelty of this study is to develop a mathematical model for the acoustic interaction force between two particles immersed in a viscoelastic fluid. The rest of the manuscript is organized as follows. First, the theoretical background is expanded by implementing a viscoelastic fluid model in the governing equations of acoustophoresis in the framework of first and second-order perturbation theory \cite{Morozov2015,Hintermuller2017,PMuller2014,Hill2014,Jakoby2017}. Then, the numerical model and boundary conditions are described. After that, the validation of this contribution is represented.
Finally, the results are presented and discussed. 
The last section concludes the letter. 

\textit{Theoretical background}. In the absence of body forces, the fluid flow governing equations in the microfluidic systems for a
viscous fluid are the kinematic continuity equation of $\rho$ and the dynamic Navier-Stokes equation
for the velocity ﬁeld $\bm v$, as
\begin{subequations}
	\begin{align}
		\partial_t{\rho}&=\mbox{\bm{$\nabla$}}\cdot[-\rho \bm{v}]\;,    \label{eq1.a}\\
		\partial_t({\rho \bm{v}})&=\bm{\nabla}\cdot[-p\textbf{I}+\bm \tau]\;.
		\label{eq1.b}
	\end{align}
	\label{eq1}
\end{subequations}
The shear stress, $\bm \tau$, of viscous fliud is defined as,
\begin{equation}
\bm\tau=\eta[\mbox{\bm{$\nabla$}}\bm
		v+(\mbox{\bm{$\nabla$}}\bm v)^\textrm{T}]+[\eta_b-\frac{2}{3}\eta](\bm\nabla\cdot \bm{v}) \textbf{I}\;,
		\label{eq2}
\end{equation}
where $\eta$ and $\eta_b$ are the dynamic and bulk viscosities, respectively. For viscoelastic
fluids, a variety of models has been reported in the literature
\cite{Morozov2015,Hintermuller2017,AJoseph1990}, to describe such fluids behavior. Here, we consider
the Maxwell model\cite{AJoseph1990} to investigate the effect of viscoelastic fluid on the acoustophoretic forces. In this model, the constitutive equation of fluid is expressed by
\begin{equation}
	\bm {\tau}+\frac{\eta}{G}\frac{\partial \bm{\tau}}{\partial t}=\eta\bm{\dot{\gamma}}\;,
	\label{eq3}
\end{equation}
where $\bm{\dot{\gamma}}$ and $G$ are the shear rate and dynamic modulus, respectively. The dot
denotes the time derivative. The symmetric $\bm{\dot{\gamma}}$ is defined as,
\begin{equation}
	\bm{\dot{\gamma}}=[\mbox{\bm{$\nabla$}}\bm
	v+(\mbox{\bm{$\nabla$}}\bm v)^\textrm{T}]\;.
	\label{eq4}
\end{equation}
Due to the frame invariance of the stress tensor, the time derivative of $\bm\tau$, is replaced with
the upper-convected time derivate as \cite{Oldroyd1950,AJoseph1990,ALandau1993}
\begin{equation}
	\mathop{\bm{\tau}}^{\nabla}=\frac{\partial \bm{\tau}}{\partial t}+\bm{ v\cdot}{\bm{\nabla \tau}}-(\bm{\nabla v})\bm{\cdot}\bm{\tau}-\bm{\tau}\bm{\cdot}(\bm{\nabla v})^\textrm{T}\;.
	 \label{eq5}
\end{equation}
Therefore, the Maxwell model is replaced with the upper-convected Maxwell model (UCM) \cite{AJoseph1990}.
In this way, the constitutive equation is represented by
\begin{equation}
\bm\tau+\lambda\mathop{\bm\tau}^{\nabla}=\eta\bm{\dot{\gamma}}\;,
	\label{eq6}
\end{equation}
where $\lambda=\eta/G$ is the relaxation time. In this contribution, the governing equations are solved using the perturbation theory \cite{ALandau1993,Lighthill2002,GorKov2019}.

\textit{First-order perturbation}. According to the perturbation theory, acoustic variables are considered as,
\begin{subequations}
	\begin{align}
		\rho(\bm r,t)&=\rho_0+\rho_1(\bm r,t)+\rho_2(\bm r,t)\;,
		\label{eq7.a}\\
		p(\bm r,t)&=p_0+p_1(\bm r,t)+p_2(\bm r,t)\;,
		\label{eq7.b}\\
		\bm v(\bm r,t)&=\bm{v}_0+\bm{v}_1(\bm r,t)+\bm{v}_2(\bm r,t)\;,
		\label{eq7.c}\\
		\bm\tau(\bm r,t)&=\bm{\tau}_0+\bm{\tau}_1(\bm r,t)+\bm{\tau}_2(\bm r,t)\;,
		\label{eq7.d}
	\end{align}
	\label{eq7}
\end{subequations}
where subscript 0, 1 and 2 indicate the zeroth, first and second order of perturbation, respectively. The fluid is considered to be at rest in the unperturbed state, so the values of $\bm v_0$ and $\bm\tau_0$ is set to be zero. First order variables are considered time-harmonic with angular frequency $\omega=2\pi f$; so that we have $\bm{v}_1(\bm{r},t)=\bm{v}_1(\bm r)e^{-i\omega t},\, \rho_1(\bm{r},t)=\rho_1(\bm r)e^{-i\omega t},\,
p_1(\bm{r},t)=p_1(\bm r)e^{-i\omega t},$ and $\tau_1(\bm{r},t)=\tau_1(\bm r)e^{-i\omega t}$. In isentropic cases, we have
$p_1=c_0^2 \rho_1$, where $c_0$ is the speed of sound \cite{Bruus2012}.

By substituting Eq. (\ref{eq7}) into the Eq. (\ref{eq1}), the first-order perturbation of the governing equations are obtained as follow,
\begin{subequations}
	\begin{align}
		i\omega \rho_1&=\rho_0 \bm{\nabla}\cdot \bm{v}_1\;,
		\label{eq8.a}\\
		-i\omega \rho_0 \bm{v}_1&=-\bm{\nabla}p_1+\bm{\nabla}\cdot \bm{\tau}_1\;.
		\label{eq8.b}
	\end{align}
	\label{eq8}
\end{subequations}
The first-order perturbation of shear stress, $\bm \tau_1$, of viscous fluid is written as,
\begin{equation}
	\bm\tau_1=\eta[\mbox{\bm{$\nabla$}}\bm
	v_1+(\mbox{\bm{$\nabla$}}\bm v_1)^\textrm{T}]+[\eta_b-\frac{2}{3}\eta](\bm\nabla\cdot \bm{v_1}) \textbf{I}\;,
	\label{eq9}
\end{equation}
and for a viscoelastic fluid, the equation (\ref{eq9}) is replaced by the following equation,
\begin{equation}
	\bm{\tau}_1=\eta^*[\bm{\nabla}\bm{v}_1+(\bm{\nabla}\bm{v}_1)^\textrm{T}]\;,
	\label{eq10}
\end{equation}
where $\eta^*=\eta_0/(1-iDe)$ denotes the complex viscosity, $\eta_0$ is the zero shear rate viscosity, and $De=\lambda\omega$ is called Deborah number. 
At $\lambda=0$, the viscoelastic fluid behaves like a viscous fluid\cite{Hintermuller2017}.

Considering all fields' time dependency as $ e^{-i\omega t}$, and inserting Eqs. (\ref{eq8.a}) and (\ref{eq9}) into the Eq. (\ref{eq8.b}), the first-order acoustic wave equation for
$p_1$ is obtained as\cite{Settnes2012},
\begin{equation}
	\nabla^2 {p_1}+\frac{\omega^2}{c_0^2}(1+i\omega\frac{\eta_b+\frac{1}{3}\eta}{c_0^2\rho_0})p_1=0\;.
	\label{eq11}
\end{equation}
For viscoelastic cases, the first-order acoustic wave equation is expressed by,
\begin{equation}
	\nabla^2p_1+\frac{\omega^2}{c_0^2}(1+i\omega\frac{2\eta^*}{c_0^2\rho_0})p_1=0\;.
	\label{eq12}
\end{equation}

\textit{Second-order perturbation}. The time-averaged second-order perturbation approximation of governing equations for a viscous fluid are expressed by,
\begin{subequations}
	\begin{align}
		\bm\nabla\cdot\left\langle\bm v_2\right\rangle+\kappa_s\left\langle\bm{v}_1\cdot\bm{\nabla}p_1\right\rangle&=0\;,
		\label{eq13.a}\\
		\bm\nabla\cdot[\left\langle\bm\tau_2\right\rangle-\left\langle p_2\right\rangle\textbf{I}-\rho_0\left\langle\bm{v_1 v_1}\right\rangle]&=0\;,
		\label{eq13.b}
	\end{align}
	\label{eq13}
\end{subequations}
where $\kappa_s=\frac{1}{\rho}(\frac{\partial\rho}{\partial p})_s=\frac{1}{\rho_0c_0^2}$ is the isentropic compressibility in classical fluid mechanics. In the adiabatic limit $\rho_1=\rho_0\kappa_sp_1$.
The time-averaged second-order perturbation of shear stress, $\bm \tau_2$, of a viscous fluid is
\begin{equation}
		\left\langle\bm\tau_2\right\rangle=\eta\left\langle\mbox{\bm{$\nabla$}}\bm
		v_2+(\mbox{\bm{$\nabla$}}\bm v_2)^\textrm{T}\right\rangle+[\eta_b-\frac{2}{3}\eta]\left\langle\bm\nabla\cdot \bm{v_2}\right\rangle \textbf{I}\;.
	\label{eq14}
\end{equation}
For a viscoelastic fluid, Eq. (\ref{eq14}) turns into
\begin{equation}
	\begin{aligned}
		\left\langle\bm{\tau}_2\right\rangle &=\eta_0\left\langle\bm{\nabla}\bm{v}_2+(\bm{\nabla}\bm{v}_2)^\textrm{T}\right\rangle\\&-\lambda\eta^*\left\langle(\bm{v}_1\cdot\bm{\nabla})[\bm{\nabla}\bm{v}_1+(\bm{\nabla}\bm{v}_1)^\textrm{T}]\right\rangle  \\&+\lambda\eta^*\left\langle(\bm{\nabla}\bm{v}_1)\cdot[\bm{\nabla}\bm{v}_1+(\bm{\nabla}\bm{v}_1)^\textrm{T}]\right\rangle\\&+\lambda\eta^*\left\langle[\bm{\nabla}\bm{v}_1+(\bm{\nabla}\bm{v}_1)^\textrm{T}]\cdot(\bm{\nabla}\bm{v}_1)^\textrm{T}\right\rangle\;.
		\label{eq15}
	\end{aligned}
\end{equation}
In the above equations the time average over full oscillation period, $T$, of each quantity, $Y(t)$, is defined
as,
\begin{equation}
		\left\langle Y \right\rangle=\frac{1}{T}\int_{0}^{T} Y(t) dt\;.
		\label{eq16}
\end{equation}
The physical, real-valued time average of two harmonically varying ﬁelds with the complex representation, is given by
\begin{subequations}
	\begin{align}
		f(\bm r,t)&=f(\bm r)e^{-i\omega t}\;,
		\label{eq17.a}\\
		g(\bm r,t)&=g(\bm r)e^{-i\omega t}\;,
		\label{eq17.b}\\
	    \left\langle fg \right\rangle&=\frac{1}{2}Re[f^*g]\;.
		\label{eq17.c}
	\end{align}
	\label{eq17}
\end{subequations}
where the asterisk denotes complex conjugation and $Re[\;]$ shows the physical real value.

\textit{Acoustic radiation force}. In this section, the acoustic radiation force applied to microparticles, immersed in viscous and
viscoelastic fluids in a standing plane wave, are studied. In our manuscript, the first-order scattering theory is
used, because the particles' size is much smaller than the incident acoustic wavelength and they act as weak
scattering points. The acoustic radiation force is obtained using the inviscid theory. This is a correct approximation for particles extremely larger than the viscous penetration depth $\delta=\sqrt{2\nu/\omega}$ ($\nu $ is the kinematic viscosity)\cite{Settnes2012}. At first, a single particle, and then a pair of particles suspended in viscous and viscoelastic
fluids in a standing plane wave (at room temperature) are studied. The incoming wave is described by the velocity
field $\bm{v}_{in}$, and the outgoing wave from the particle is described by the velocity field $\bm{v}_{sc}$.
Therefore, the first-order velocity field $\bm{v}_1$ is expressed by
\begin{equation}
	\bm{v}_1=\bm{v}_{in}+\bm{v}_{sc}\;.
	\label{eq18}
\end{equation}
The acoustic radiation force acting on each particle is obtained by
\begin{equation}
	\begin{aligned}
		 \bm{F}_{rad}&=-\int_{\partial\Omega}da (\left\langle {p_2}\right\rangle\bm n+\rho_0\left\langle{(\bm{n}\cdot \bm{v_1})\bm{v_1}}\right\rangle)
		 \\&=-\int_{\partial\Omega}da ([\frac{\kappa_s}{2}\left\langle{p_1^2}\right\rangle-\frac{\rho_0}{2}\left\langle{v_1^2}\right\rangle]\bm{n}+\rho_0\left\langle{(\bm{n}\bm{\cdot {v_1}})\bm{v_1}}\right\rangle)\;,
		\label{eq19}
	\end{aligned}
\end{equation}
where $\partial\Omega$ is the particle's surface, and $\bm n$ is the unit-vector normal to the surface.

\begin{figure}[ht!]
	\begin{center}
			{\includegraphics[width=80mm]{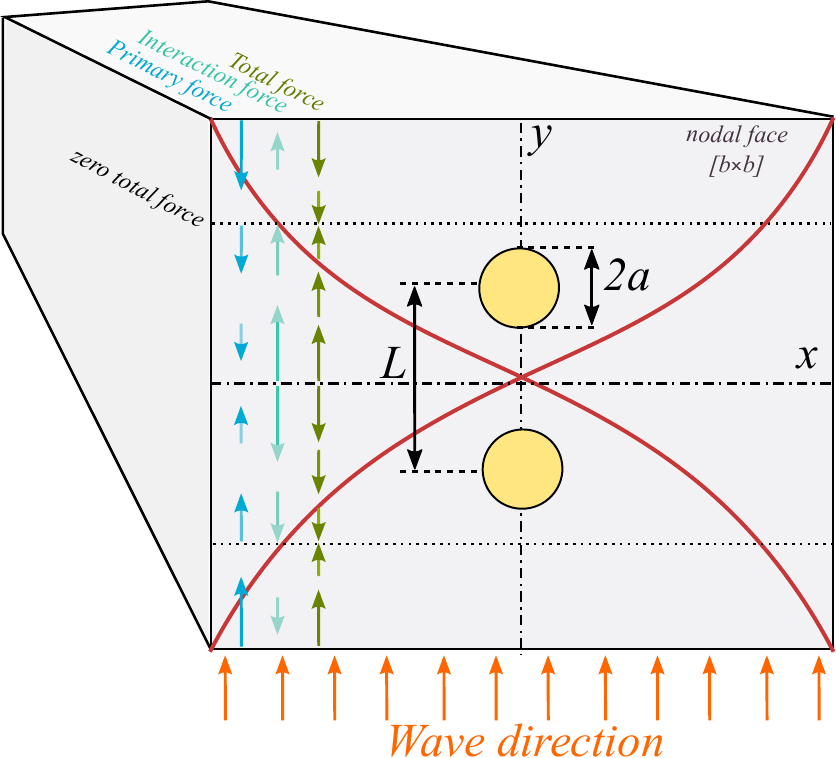}}	
	\end{center}
	\caption{Configuration of a pair of spheres in a microchannel. $L$ is the center-to-center distance between two particles with a radius of $a$.}
	\label{fig1}
\end{figure}

\textit{Numerical model and boundary conditions}. As shown in Fig. \ref{fig1}, a microchannel with a square cross-section of the size, $b=500\; \mu m$ in the XY plane is considered. The incident wave is a resonant pressure wave of the form $p=p_{amp}\; sin(2\pi y/w)$, in which $w$ and $p_{amp}$ are the wavelength and the amplitude of the wave. In our simulation, the sphere's diameter is considered $8\;\mu m$. 

\begin{figure*}[ht]
	\begin{center}
	\includegraphics[width=180mm]{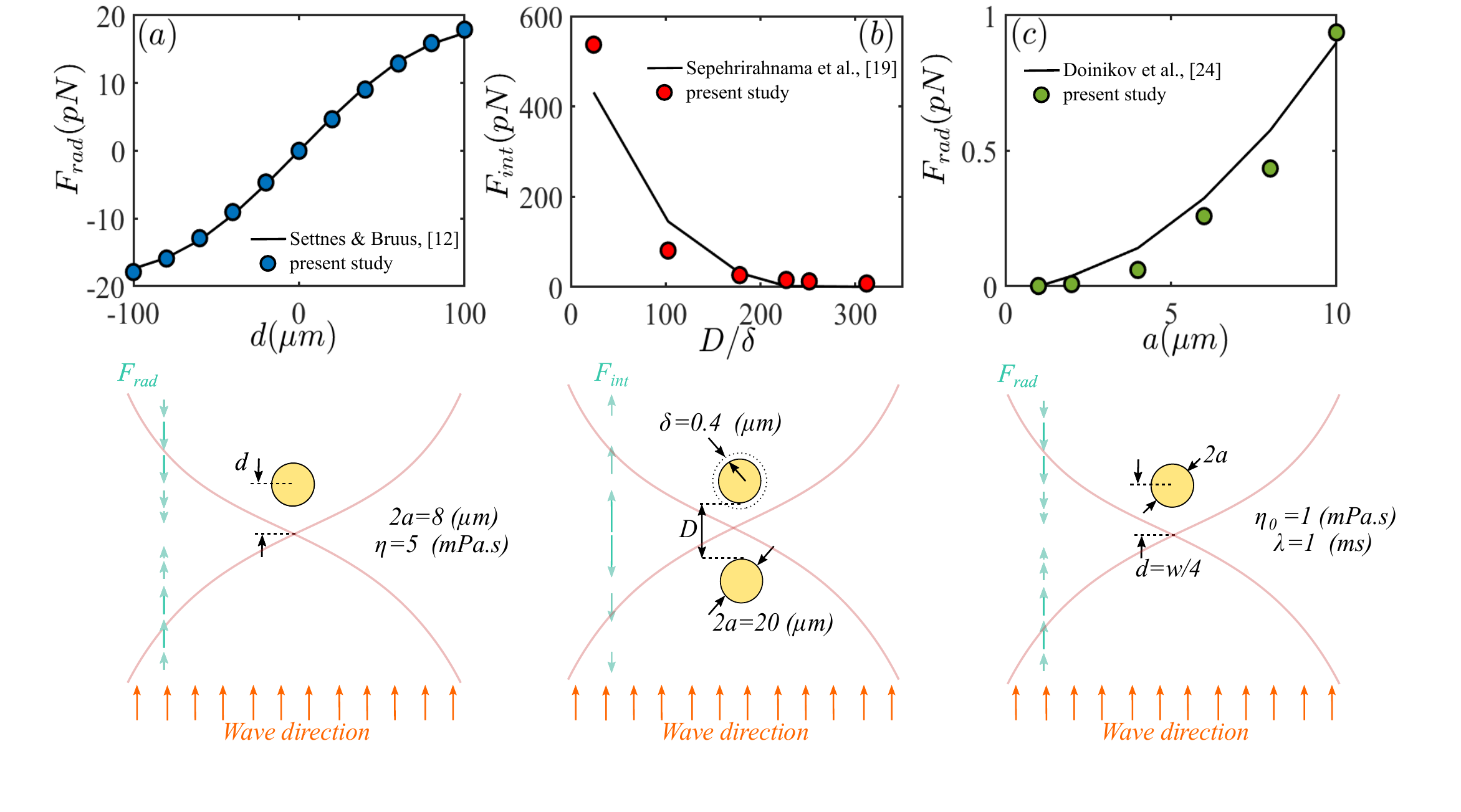}
	\end{center} \vspace{-0.5 in}
	\caption{(a) Acoustic radiation force on a sphere with $a=4\;\mu m$ in a viscous fluid. (b) Acoustic interaction force between a pair of spheres with $a=10\;\mu m$ in a viscous fluid. (c) Acoustic radiation force on a  polystyrene particle in a viscoelastic fluid.}
	\label{fig2}
\end{figure*}

The governing equations are solved for one- and two-particle systems of spheres suspended in viscous and viscoelastic fluids using the finite element method. The weak-form-PDE is used to solve these
equations \cite{Elnaz2019,Elnaz2021}. In the weak-form-PDE method, first, the flow equations are written as source-free flux formulation,
$\bm\nabla\cdot\bm J+ F=0$. Then, they are converted to the weak-form. At the end, the weak-form equations are
solved by finite element method.

Considering $\bm{J}$ as a vector and $F$ as a scalar (e.g. equations \ref{eq8.a} and \ref{eq13.a}), the weak form of a
source-free flux equation is given by
\begin{equation}
	\int_{\Omega}\left[-\bm\nabla\widetilde{\psi}\cdot
	\bm{J}+\widetilde{\psi}F\right]d\bm r=0\;,
	 \label{eq20}
\end{equation}
where $\widetilde{\psi}$ is a test function. For $\bm{J}$ as a tensor and $\bm{F}$ as a vector (e.g. equations \ref{eq8.b} and \ref{eq13.b}), the weak form of a source-free flux equation is
\begin{equation}
	\int_{\Omega}\left[-\bm\nabla\widetilde{\Psi}_m\cdot
\bm{J}+\widetilde{\Psi}_m\cdot\bm F\right]d\bm r=0\;, \;\;\textrm{for all $m$}
	\label{eq21}
\end{equation}
where $\widetilde{\Psi}_m$ is the $m$-th component of a test vector $\bm \Psi$. In all cases, the zero-flux boundary condition, $\bm J \cdot \bm{n} =0$, is considered. All boundaries of Fig. \ref{fig1} including microchannel and particles, except the bottom wall, are considered as hard walls. Therefore, the velocity field for these boundaries is $\bm v \cdot \bm n =0$. For the bottom wall the boundary condition is $p=p_{amp}\;sin(2\pi y/w)$.

The maximum mesh size on particles' boundaries is equal to $0.01\mu m$, and $2\mu m$ in bulk region. The mesh element growth rates on particles' boundaries and in bulk are $1.0003$ and $1.1$, respectively.
The fluid inside the microchannel is supposed to be a quiescent viscous and viscoelastic fluid. The physical parameters of fluids and the incident acoustic wave at temperature $T=25^{\circ}\mathrm{C}$ and pressure $p_0=0.1013 MPa$ are presented in Table \ref{TB1}.

\begin{table}
	\caption{Physical parameters of fluids and incident acoustic wave at $T=25^{\circ}\mathrm{C}$ and
		$p_0=0.1013 {\;\rm  MPa}$. $\eta_0=16.9\;mPa.s$ and $\lambda=7.8\;ms$ are the real experimental data\cite{Settnes2012,Brust2013}.}
	\label{TB1}
	\begin{center}
		\begin{tabular}{lccc}
			\hline\hline
			Parameter&Symbol&Value&Unit\\
			\hline
			Wave amplitude & $p_{amp}$ & 1 & bar\\
			Wave length & $w$ & 1000 & $\mu$m\\
			Actuation frequency & $f$ & 1.5 & MHz\\
			Mass density & $\rho_0$ & 1000 & kg/m$^3$\\
			Speed of sound & $c_0$ & 1500 & m/s\\
			Bulk viscosity & $\eta_b$ & 2.485$\times10^{-3}$  & Pa$\cdot$ s\\
			Zero shear rate viscosity & $\eta_0$ & 5,\;9,\;16.9,\;25  & Pa$\cdot$ s\\
			Relaxation time & $\lambda$ & 0,\;0.6,\;4,\;7.8,\;16 &  ms\\
			\hline
		\end{tabular}
	\end{center}
\end{table}
By solving the governing equations, the pressure and velocity fields are obtained. Then the acoustic radiation forces are calculated by Eq. (\ref{eq19}).

\textit{Validation}. In order to verify our using method in this contribution, the obtained results for the case of one-particle system surrounded by viscous fluid are compared with those reported by Settnes and Bruus \cite{Settnes2012}. For this purpose, a $4-\mu m$ solid sphere ($ka=0.025<<1$, $k=2\pi/w$ is the wave number) is considered with its particle-distance $d$ from the pressure node. The host fluid has a viscosity of $\eta=5\;mPa.s$, density of $\rho_0=1000\;kg/m^3$, and the speed of sound $c_0=1500\;m/s$. The frequency of the standing wave is $1.5 MHz$ and its wavelength is $1000\;\mu m$. The pressure amplitude is $1\; bar$.

Figure $\ref{fig2}(a)$ shows the acoustic radiation force acting on the sphere for various $d$. It can be seen that our simulation results are in very good agreement with Settnes and Bruus results, which actually confirms the correctness of our data.

Besides, we have considered a two-particle system including two $10-\mu m$ solid spheres with their surface-to-surface distance $D$ in a viscous fluid with $\delta=0.4\;\mu m$ and calculated the acoustic interaction force between them. Then, we compared our data with those calculated by Sepehrirahnama et al. \cite{Sepehrirahnama2016}.  The other parameters are as in Fig. $\ref{fig2}(a)$. Our obtained results presented in Fig. $\ref{fig2}(b)$ are in agreement with the data by Sepehrirahnama et al..   

Moreover, we have also made a comparison between our results and those of Doinikov et al.\cite{Dual2021} for the case of one-particle system including a polystyrene particle located halfway between the pressure node and the antinode in a viscoelastic fluid with $\eta_0=1 Pa.s$ and $\lambda=1 ms$. The parameters of the particle material are the following: the density $\rho_s=1050\;kg/m^3$, Young's modulus $E=0.32\times10^{10} Pa$, and Poisson's ratio $\nu=0.35$. The frequency of the standing plane wave is $1\;MHz$ and the pressure amplitude is $10\; kPa$. The other parameters are as in Fig. $\ref{fig2}(a)$. 
In Fig. $\ref{fig2}(c)$, the vertical axis shows the acoustic radiaton force acting on the particle.
The results are shown for different particle's radius $a$. The results are compatible with their. 

\textit{Results and discussion}. In this contribution, extensive numerical calculations are carried out to study the acoustic radiation forces applied to spherical particles in a two-particle system. The particles suspended in viscous and viscoelastic fluids are exposed to an external standing plane wave. The obtained results indicate that the fluid's rheological parameters like viscosity and relaxation time play important roles on the magnitude of acoustic forces experienced by particles. In the following, results and their implications are discussed.

\textit{Total and primary forces}. For a two-particle microfluidic system, the total radiation force applied to each particle is divided into the primary and secondary (interaction) forces. The primary force applied to each particle is due to the incoming waves on and scattered waves from the target particle. The secondary force is caused by the scattered waves from the other particle on the target one. First, a two-particle microfluidic system filled with a viscous fluid is considered.

\begin{figure}[ht!]
	\begin{center}
			\includegraphics[width=80mm]{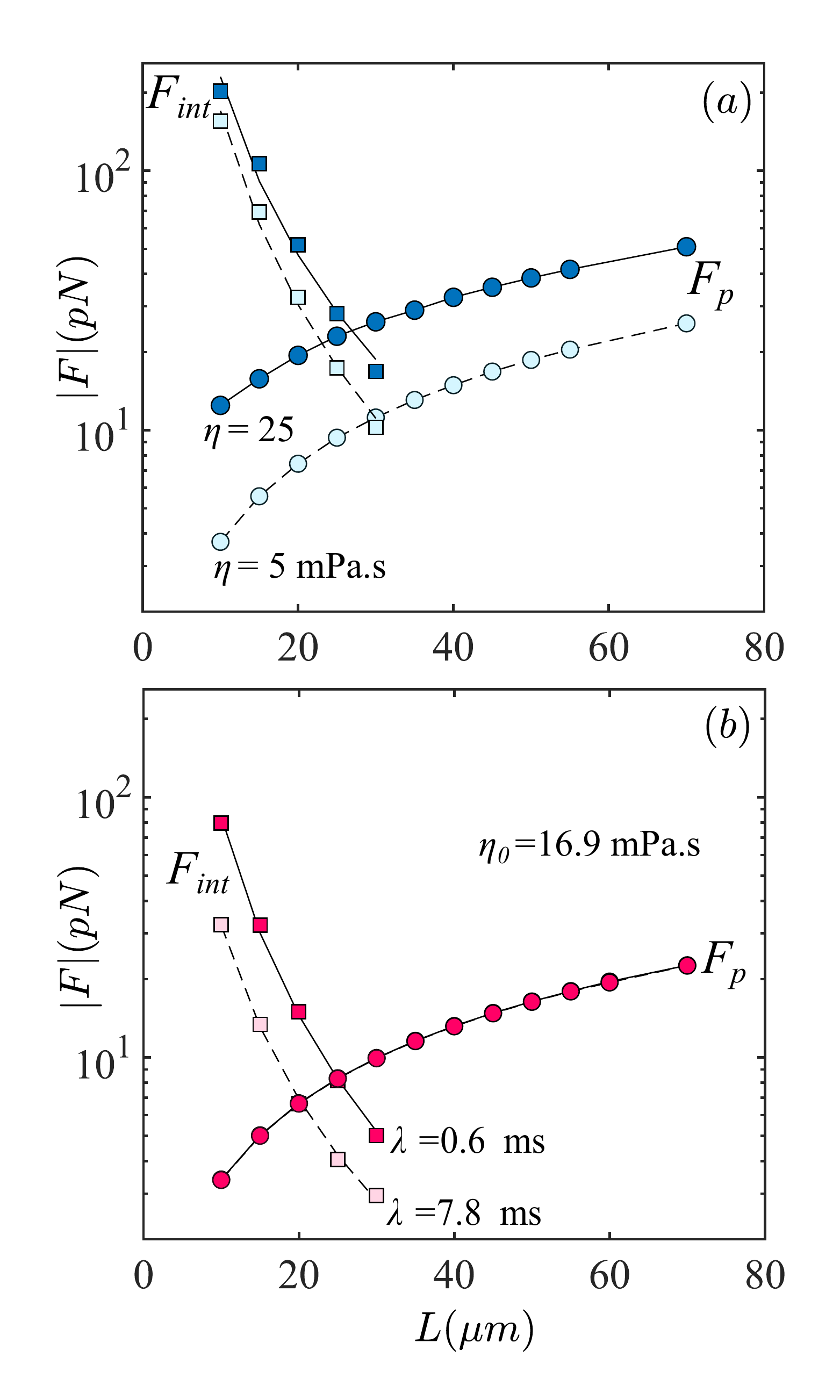}\label{fig3.a}
			
	\end{center}\vspace{-0.3 in}
	\caption{Primary ($\circ$) and secondary ($\square$) forces applied to each particle in a two-particle system for (a) viscous fluids with $\eta = 5$ ({\color{black}$--$}) and $25\;mPa.s$ ({\color{black}$-$}), and (b) viscoelastic fluids with $\eta_0=16.9\;mPa.s$, $\lambda=0.6$  ({\color{black}$-$}) and $7.8\;ms$ ({\color{black}$--$}).}
	\label{fig3}
\end{figure}

In Fig. $\ref{fig3}(a)$, the magnitude of the primary and secondary forces has been plotted in the semi-logarithmic scale for two different values of viscosity. For small $L$ values, the secondary force between two objects is strong. In contrast, the primary force is close to zero. For large distances between the particles, when the spheres are far from the pressure node, the primary force is dominant, while the interaction between the particles becomes noticeably weak. For an intermediate value of $L$, these two forces neutralize each other. It can be seen that for a definite $L$ value, forces increase with fluid viscosity. Furthermore, it is noticeable that the effect of viscosity on the primary force is more than on the secondary force. 

In the next step, the primary and secondary forces for viscoelastic fluids with various $\lambda$ values are presented in Fig. $\ref{fig3}(b)$. It is clear that the value of the relaxation time play an important role on the magnitude of the interaction force. Besides, the force decreases by $\lambda$. 

\begin{figure}[ht!]
	\begin{center}
			\includegraphics[width=80mm]{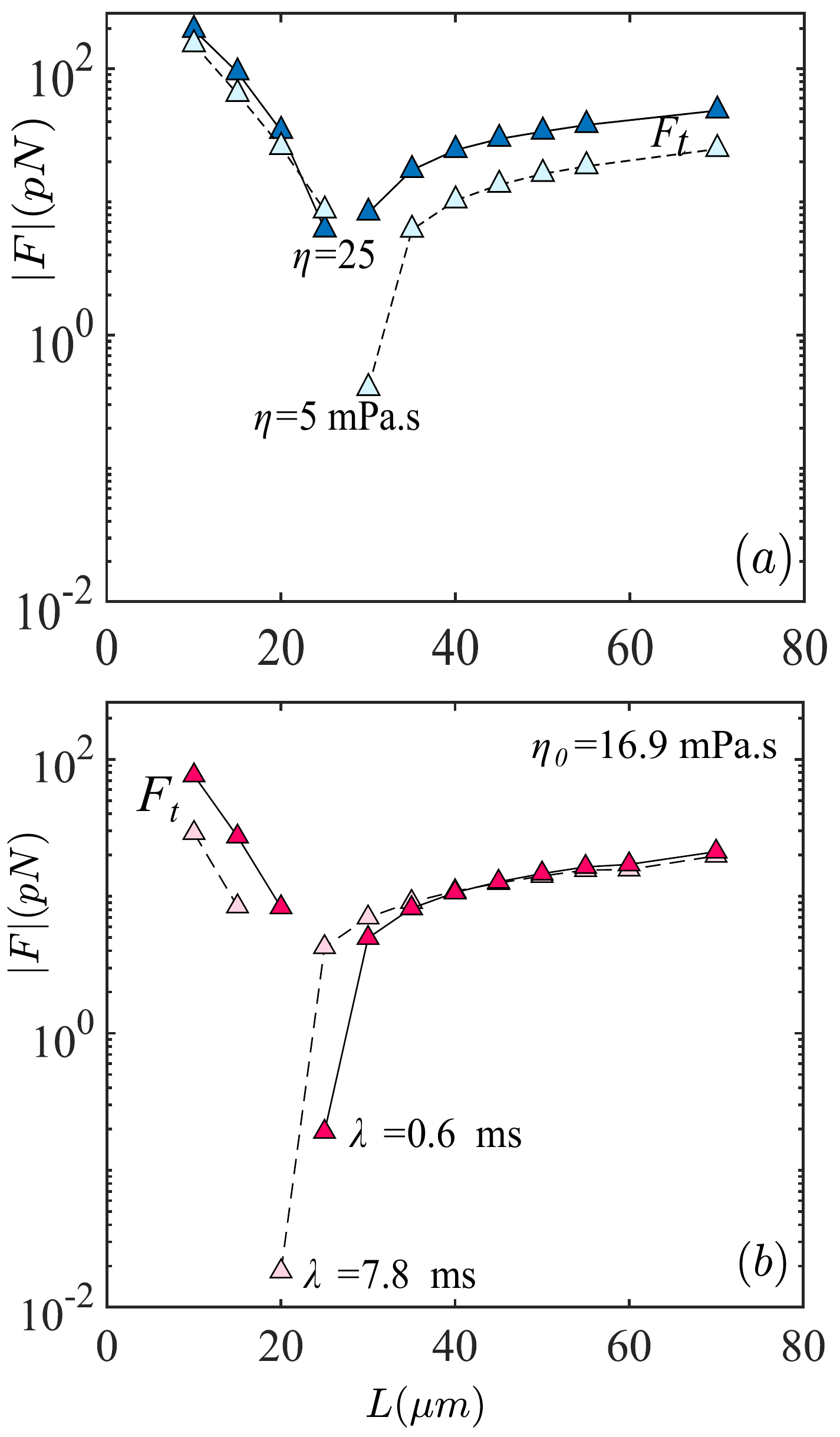}\label{fig4.a}
		
	\end{center}\vspace{-0.3 in}
	\caption{Total radiation force ($\bm{F_t}=\bm{F_p}+\bm{F_{int}}$) applied to each particle in a two-particle system for (a) viscous fluids with $\eta = 5$ ({\color{black}$--$}) and $25\;mPa.s$ ({\color{black}$-$}), and (b) viscoelastic fluids with $\eta_0 =16.9\;mPa.s$, $\lambda=0.6$  ({\color{black}$-$}) and $7.8\;ms$ ({\color{black}$--$}).}
	\label{fig4}
\end{figure}

In Fig. $\ref{fig4}(a)$, the magnitude of the total radiation force has been shown for different values of $\eta$. It can be seen that the total radiation force is large at small distances which is due to the large interaction force between particles in small $L$ values. The total force has a discontinuity in the graph and this is because of the zero value of the force, since it cannot be shown on the semi-logarithmic scale. This is where the primary and secondary forces balance each other. By increasing the inter-particle distance, the interaction force is negligible, so the total and primary forces are almost equal. 

The total force for viscoelastic fluids with various $\lambda$ values are presented in Fig. $\ref{fig4}(b)$. The zero value of total force occurs in smaller $L$ for larger $\lambda$ values, in compatible with the relation $\eta^*=\eta_0/(1-i\lambda\omega)$ and the results of Fig. $\ref{fig4}(a)$.

The acoustic interaction force between spheres is examined
in detail in the following.

\begin{figure}[ht!]
	\begin{center}
			{\includegraphics[width=80mm]{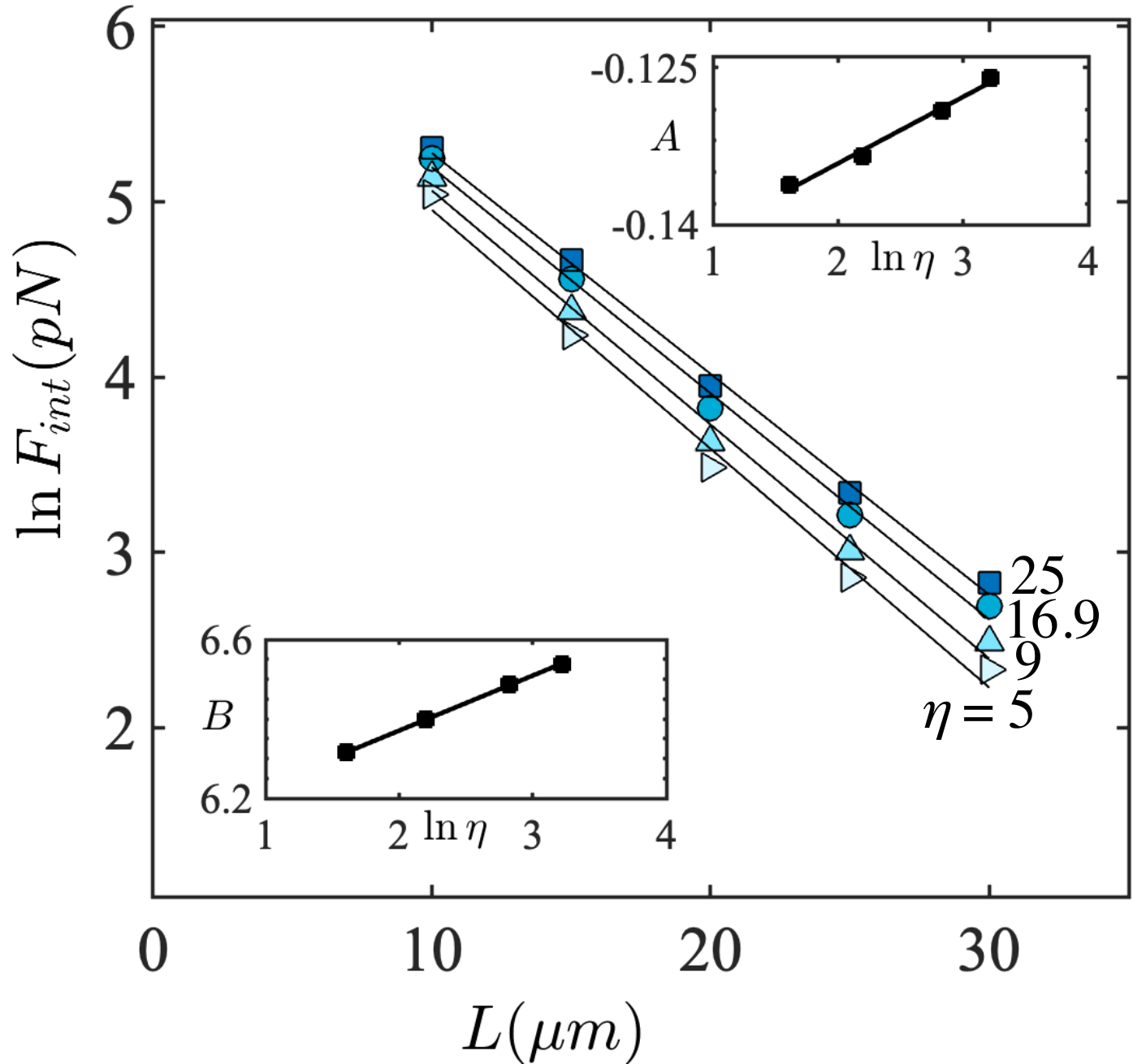}}	
	\end{center}\vspace{-0.3 in}
	\caption{Acoustic interaction force between two spheres for viscous fluids with $\eta=5$ ($\blacktriangleright$), $9$ ($\blacktriangle$), $16.9$ ($\bullet$) and $25$ ($\blacksquare$) $mPa.s$. Insets are the slopes (A) and intercepts (B) of $\ln F(L, \eta)$.}
	\label{fig5}
\end{figure}

\textit{Acoustic interaction force}. Figure $\ref{fig5}$ represents the acoustic interaction force between two spheres suspended in a viscous fluid as a function of fluid viscosity $\eta$ and inter--particle distance $L$. 
The data show that the value of $F$ exponentialy decreases by $L$ as $\ln F = A(\eta) L + B(\eta)$, compatible with the results shown in  Fig. \ref{fig2}(b). The functionality of $A$ and $B$ are presented as insets. According to these data one may write the interaction forces between two spheres as
\begin{equation}
	F(L,\eta)=\exp{[(0.006\ln{\eta}-0.15)L+(0.14\ln{\eta}+6.09)]}\;.
	\label{eq22}
\end{equation}
Equation (\ref{eq22}) indicates for a definite inter--particle distance, $F$ increases by $\eta$.

\begin{figure*}[ht!]
	\begin{center}
		\begin{tabular}{cc}
		
			{\includegraphics[width=190mm]{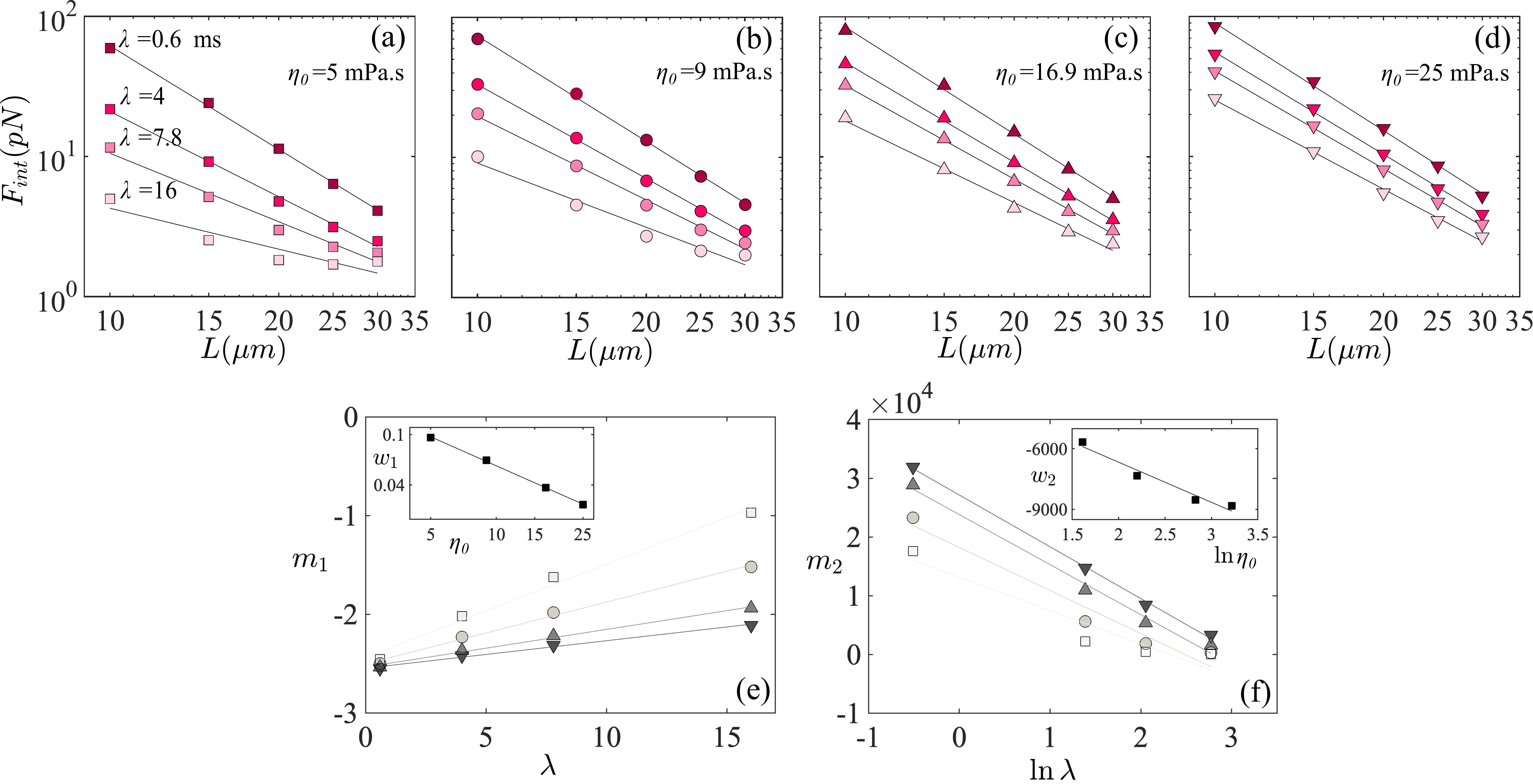}}
			
		\end{tabular}
	\end{center}\vspace{-0.3 in}
	\caption{Acoustic interaction force between a pair of spheres versus $L$ for viscoelastic fluid with viscosities (a) $\eta_0 =5$, (b) $9$, (c) $16.9$, and (d) $25$ $mPa.s$. The data are for $\lambda=0.6$ ({\color[rgb]{0.67,0,0.27}$\blacksquare$}), $4$ ({\color[rgb]{1,0,0.4}$\blacksquare$}), $7.8$ ({\color[rgb]{1,0.5,0.7}$\blacksquare$}) and $16$ ({\color[rgb]{1,0.84,0.9}$\blacksquare$}) $ms$. (e) The slopes and (f) intercepts of $F(L,\eta,\lambda)$ for viscosities $\eta=5$ ($\blacksquare$), $9$ ($\bullet$), $16.9$ ({\color{black}$\blacktriangle$}), $25$ ({\color{black}$\blacktriangledown$}) $mPa.s$. Insets show slopes of $m_1$ and $m_2$.}
	\label{fig6}
\end{figure*}

In the next step, we obtain the acoustic interaction force between a pair of spheres suspended in a viscoelastic fluid. The interaction force between two particles as a function of $L$ for various $\eta_0$ and $\lambda$ is plotted in Figs. \ref{fig6}(a) -- \ref{fig6}(d). The results show that for definite $\lambda$ values, $F$ behaves as a decreasing power-law function of $L$. Therefore, we can consider the interaction force as $F=m_2 L^{m_1}$. The functionality of $m_1$ and $m_2$ versus $\lambda$ for various $\eta_0$ are plotted in Figs. \ref{fig6}(e) and \ref{fig6}(f), respectively. Insets are the slopes of $m_1$ and $m_2$ as functions of $\eta_0$. By fitting all these data the interaction force
is obtained as,
\begin{equation}
   \begin{aligned}
	F(L,\eta_0,\lambda)=&[(-1986.5\ln\eta_0-2695.5)\ln\lambda+8733.7\ln\eta_0\\&-927.49]
	 \times L^{(0.76\lambda{\eta_0}^{-0.33}-0.07\ln{\eta_0}-2.34)}\;.
	\label{eq23}
	\end{aligned}
\end{equation}
Equation (\ref{eq23}) shows that $F$ increases  by fluid's viscosity and decreases by its relaxation time. This mathematical formula shows the behavior of the interaction force between two spherical particles located close to each other immersed in a viscoelastic fluid that no one has obtained until now.

\textit{Conclusion}. 
In this letter, it could be viewed as the first simulational modeling of the acoustic interaction force between two spherical particles immersed in a viscoelastic fluid was developed. We have used the first- and second-order perturbation theories for the governing equations to develop the mathematical model based on the Upper-Convected Maxwell equation. 

In order to validate the developed mathematical model, firstly we compared it with the literature. We initially verified it for the case of one particle system surrounded by a viscous fluid with the results obtained by Settnes and Bruus\cite{Settnes2012}. Then, for the case of a two-particle system in a viscous fluid, the results were compatible with those calculated by Sepehrirahnama et al.\cite{Sepehrirahnama2016}. 
Finally, we made a comparison between our results and those of Doinikov et al.\cite{Dual2021} for the case of a one-particle system in a viscoelastic fluid. The results were in agreement with those obtained by the literature. 
Indeed, in the presented model, the particles' radius should be much smaller than the acoustic wavelength and bigger than the viscous penetration depth.

We concluded that the interaction force between the spherical particles is larger for viscous fluids compared to viscoelastic ones at the same zero shear rate viscosity values. This occurred due to the relaxation time effect. In fact, this letter presents the dependence of interaction force on the rheological parameters of the fluid media in which the particles are immersed. In addition, we showed that the interaction force increases by the viscosity and reduces by the relaxation time. A decreasing exponential function of inter-particle distance was obtained, by fitting simulation data, for the interaction force between particles suspended in the viscous fluid. However, a power-law function was obtained in the case of a viscoelastic one.
In future work, the effect of drag force as a function of particles' size induced by acoustic streaming, and particle concentration will be considered in a viscoelastic media.


\end{document}